\documentclass[sn-vancouver]{sn-jnl}% Vancouver Reference Style
\jyear{2024}%

\usepackage{amsmath}
\usepackage{graphicx}
\usepackage{booktabs}
\usepackage{float}
\usepackage[export]{adjustbox}
\usepackage{capt-of}
\usepackage{subcaption}

\usepackage{xcolor}
\newcommand{\change}[1]{\textcolor{black}{#1}}

\title[Social Influence on Vaccination Decisions]{\change{Perceived Social Influence on Vaccination Decisions: A COVID-19 Case Study}}

\author[1]{\fnm{Denise} \sur{Yewell}}\email{dyewell@vols.utk.edu}

\author[2]{\fnm{R. Alexander} \sur{Bentley}}\email{rabentley@utk.edu}

\author*[1]{\fnm{Benjamin D.} \sur{Horne}}\email{bhorne6@utk.edu}

\affil[1]{\orgdiv{School of Information Sciences}, \orgname{University of Tennessee, Knoxville}, \orgaddress{\city{Knoxville},  \state{TN}, \country{USA}}}

\affil[2]{\orgdiv{Anthropology Department}, \orgname{University of Tennessee, Knoxville}, \orgaddress{\city{Knoxville},  \state{TN}, \country{USA}}}

\abstract{In this study, we examine the perceived influence of others, across both strong and weak social ties, on COVID-19 vaccination decisions in the United States. We add context to social influence by measuring related concepts, such as perceived agreement of others and perceived danger of COVID-19 to others. We find that vaccinated populations perceived more influence from their social circles than unvaccinated populations. This finding holds true across various social groups, including family, close friends, and neighbors. Vaccinated participants perceived that others agreed with their decision to get vaccinated more than unvaccinated participants perceived others to agree with their decision to not get vaccinated. Despite the clear differences in perceived social influence and agreement across the groups, the majority of participants across both vaccinated and unvaccinated populations perceived no social influence from all social group in their decisions. Aligning with this result, we find through open-ended responses that both vaccinated and unvaccinated participants frequently cited fear as a motivating factor in their decision, rather than social influence: vaccinated participants feared COVID-19, while unvaccinated participants feared the vaccine itself.}


\begin{document}
\maketitle

\section{Introduction}

One thing learned from COVID-19 is that the spread of a pandemic in nations around the world depended on not only on accurate epidemiological information and government responses \cite{Prather_etal_2020, Aksoy_etal_2020, Hale_etal_2020, Bedford_etal_2020, Frey_etal_2020, Munster_etal_2020,  Chowell&Mizumoto_2020, Zhang&Qian_2020} but also the varied behaviors of individuals, groups, and cultures \cite{Maharaj&Kleczkowski_2012, Funk_etal_2010, Guiteras_etal_2015, Zhang&Centola_2019, gharzai2020playing, Ruck_etal_2021a}. Since culture is the context for behavior \cite{Zhang&Centola_2019, Gelfand_etal_2021}, and cultural values vary substantially across the world \cite{Aksoy_etal_2020, Inglehart&Welzel_2005, Ruck_etal_2020}, the effect of COVID-19 in terms of cases and deaths, was affected not only by government actions but by cultural values and social norms in their populations \cite{Ruck_etal_2021a, moehring2021surfacing}.  

These varied behaviors have been well-studied, particularly early in the COVID-19 pandemic, before vaccines were widely available. These studies have used socioeconomic and public health variables to explain COVID-19 variation within the United States \cite{Desmet&Wacziarg_2020, Ruck_etal_2021b} and also globally \cite{deOliveira_etal_2020}. In the context of the literature on the effect of socioeconomic factors on COVID-19 \cite{Aksoy_etal_2020, Hale_etal_2020, Bedford_etal_2020, Frey_etal_2020, Munster_etal_2020, Chowell&Mizumoto_2020, Zhang&Qian_2020, Maharaj&Kleczkowski_2012, Funk_etal_2010, Guiteras_etal_2015, Zhang&Centola_2019}, cultural effects have been examined in concert with known risks such as obesity and advanced age, together with variables describing government efficiency and public trust in institutions \cite{Ruck_etal_2021a, Ruck_etal_2021b}.  

In terms of the behavioral dynamics of vaccination, those who are not vaccinated may feel less urgency due to a lowered perceived risk of infection \cite{verelst2016, fineberg2013paradox, jansen2003measles}. Furthermore, even those infected with a coronavirus or flu may not exhibit obvious symptoms to others \cite{Li_etal_2020}. Inadequate testing of populations \cite{Munster_etal_2020} can mean that the majority of infections in a population are undocumented \cite{Li_etal_2020, Munster_etal_2020, Bendavid_etal_2020, Baud_etal_2020, Marchant_etal_2020}.  This can lead individuals to underestimate the risk of a virus, especially in context with other socio-economic concerns \cite {Mani_etal_2013, Trueblood_etal_2020}.  Social influence from peers to vaccinate may be weak because\textemdash unlike visible protections such as mask-wearing or conspicuous lack of people in public spaces\textemdash  vaccination is often comparatively less visible. In this situation, the perceived benefits of vaccination may be outweighed by economic and/or psychological costs \cite{Trueblood_etal_2020, Maharaj&Kleczkowski_2012, Pritchard_etal_2022a, greenstone2020does}.
 
The complexity of the levels of voluntary vaccination lies in the multiple drivers of behavioral change, including information and social learning \cite{Carrignon_etal_2022, Funk_etal_2010, Guiteras_etal_2015,Christakis&Fowler_2013, Bentley_etal_2011, Silk_etal_2022}. Ideally, decisions would be determined by their intrinsic payoffs within their socio-ecological environment \cite{Gibson&Lawson_2015}. In the real world, decisions are made by people who combine observational learning, which produces noisy information, and social learning, which diffuses that information to others \cite{Hoppitt&Laland_2013, Rendell_etal_2011}. The transparency of learning is a crucial parameter \cite{Hauser_etal_2014}; the less transparent the payoffs are, the more noise and heterogeneity enters into the behavioral dynamics \cite{Li&Lee_2009, Caiado_etal_2016, Vidiella_etal_2022}. While humans are motivated to avoid cues associated with pathogens, through emotions such as fear or disgust \cite{Schrock_etal_2020}, COVID-19 and similar viruses displayed few cues in asymptomatic cases \cite{kronbichler2020asymptomatic}.  

These considerations suggest that drivers of social distancing include two dynamic factors: observable risks and observable behaviors. In everyday life, we may expect more intimate social influence, such as advice from parents or other close family members, to be the strongest for vaccinations. Vaccination can be made more visible of course through public outreach, but its acceptance will vary based on group norms and potentially large differences in risk perception due to misinformation. Acceptance may also vary based on a range of cultural memory effects, such as the past misuse of vaccination on marginalized groups. In contrast, if genuine benefits of vaccination become more transparent to the public, individual cost-benefit decisions can support the behavior spreading through social influence \cite{Carrignon_etal_2022, Guiteras_etal_2015, Silk_etal_2021, Pritchard_etal_2022b, Banerjee_etal_2013, bentley2024cultural}.

Here we examine one under-researched facet of this complex problem: the perceived social influence of others on vaccination decisions. While prior work has shown that social influence matters in an array of health-related decisions, including social distancing and mask wearing during the COVID-19 pandemic \cite{tunccgencc2021social}, examining social influence on COVID-19 vaccine decisions has had comparably less study, particularly across strong and weak social ties. Hence, our work seeks to fill this gap and add robustness to the current findings in the literature. More broadly, further work in this area adds to our understanding of the interactions between socio-cultural dynamics and health-related decision making. Our \change{three} research questions are as follows: 
\begin{itemize}
    \item \textbf{RQ1:} Does the perception of social influence on vaccination decisions change across vaccinated and unvaccinated populations?
    \item \textbf{RQ2:} Does the perception of social agreement with vaccination decisions change across vaccinated and unvaccinated populations?
    \item \textbf{RQ3:} Does the perceived danger of COVID-19 to others change across vaccinated and unvaccinated populations?
    %\item \textbf{RQ4:} If not by social influence, what other reasons do vaccinated and unvaccinated populations think influence their vaccination decision?
\end{itemize}

\change{From our quantitative analysis, we find that vaccinated people perceived significantly more social influence from their families, close friends, and neighbors in their decision to get vaccinated than unvaccianted people perceived in their decision to not get vaccinated. However, notably, the majority of participants across both vaccinated and unvaccinated populations perceived no social influence from all social group. Likewise, vaccinated people perceived significantly more agreement with their decision to get vaccinated from their families, close friends, coworkers, and neighbors than unvaccinated people perceived with decision to not get vaccinated. Surprisingly, perceived danger of COVID-19 to varying social circles did not significantly differ across vaccinated and unvaccinated populations. Yet, through open-ended responses that both vaccinated and unvaccinated participants frequently cited fear as a motivating factor in their decision, rather than social influence or agreement: vaccinated people feared COVID-19, while unvaccinated people feared the vaccine itself.}

\section{Methods}\label{sec:qualmethods}

\change{\subsection{Survey}}
To answer our research questions, we designed and deployed a survey on the online survey platform Prolific (\url{prolific.co}). The survey contained questions on perceived influence of varying social circles, perceived agreement of varying social circles, perceived danger of COVID-19 to varying social circles, and an open-ended reasoning question. 

Specifically, we asked participants ``How much influence did each social group have on your decision to get vaccinated?'' or ``
How much influence did each social group have on your decision to not get vaccinated?'' as a matrix Likert Scale question across five groups: \textit{family members}, \textit{close friends}, \textit{co-workers}, \textit{church or social club}, and \textit{neighbors or community members}. Participants could answer on a 5-point scale: \textit{none at all}, \textit{a little}, \textit{a moderate amount}, \textit{a lot}, or \textit{a great deal}. Moreover, we asked participants ``How much do you think each social group agrees with your decision to get vaccinated?'' or ``How much do you think each social group agrees with your decision to not get vaccinated?'' as a matrix Likert Scale question across the same five groups. Participants could answer on a 3-point scale: \textit{disagree}, \textit{do not care}, or \textit{agree}. We also asked participants ``How likely do you think members of each social group are to get severe COVID-19 or die from COVID-19?'' as a matrix Likert Scale question across the same five groups. Participants could answer on a 5-point scale: \textit{extremely unlikely}, \textit{somewhat unlikely}, \textit{neither likely nor unlikely}, \textit{somewhat likely}, or \textit{extremely likely}.

To add context to these core questions, we wanted to understand both the size of participants social circles and how much they trust members of their social circles. Hence, we asked participants ``To the best of your ability, type in the first names or relationship titles (mother, best friend, cousin, co-worker, etc.) of each person you voluntarily had a conversation with in the past week.'' and ``Of those people you listed in the previous question, how many of them would you turn to for advice with a major personal problem?'' to approximate the number of people they interact with on a regular basis and how much they trust those individuals. This line of questioning is similar to the questions asked in prior work to approximate the size of one's social circle \cite{dunbar1995social, tunccgencc2021social}.

Lastly, we asked ``In your own words, why did you get a COVID-19 vaccine?'' or ``In your own words, why did you not get a COVID-19 vaccine?'' as an open-ended question to elicit reasoning outside of our social influence questions. Our goal with this question was to add further context, as participants' primary reasoning for getting or not getting vaccinated may not be based on social cues.

In addition, we collected several types of demographic information for comparison to previous studies on COVID-19 vaccination and related behaviors. These demographic questions included: gender, race, political leaning, income, education, rural-urban residency, and rural-urban identity (as defined by \cite{lunz2022rural}). Details on each question and the scales used can be found in the supplemental material.
 
\change{\paragraph{Deployment}}
On May 1st 2021, we deployed a pilot survey for 15 participants to measure if our estimated survey time and payment amount were correct. There were no issues found during the pilot, so the same payment parameters and questions were used in two more batches of survey deployment on May 2nd 2021. The pilot survey was given to any Prolific user in the United States, whether they had been vaccinated for COVID-19 or not. However, to ensure balance of vaccinated and unvaccinated participants, we deployed the actual survey to two pre-screened groups provided by Prolific: participants who were vaccinated for COVID-19 and participants who were unvaccinated for COVID-19. Using these two panels of participants, we deploy the survey for approximately 500 participants in each group. According to Prolific, 1,306 participants were eligible to take our survey in the unvaccinated U.S. group and 8,569 participants were eligible to take our survey in the vaccinated U.S. group. 

The median time per participant in the pilot study was just over 3 minutes, in the unvaccinated group study it was 4 minutes and in the vaccinated group study it was 3 minutes. All participants lived in the United States and were paid \$1.00 for survey completion. This survey was approved by The University of Tennessee's IRB.

\change{\subsection{Analysis}}\label{sec:qualmethods}
We analyzed the data from both the pilot deployment and actual deployment, totalling to 1015 responses. After filtering out responses that did not pass our attention check question, we had 1000 survey responses to analyze between the three deployments, with 486 unvaccinated participants and 514 vaccinated participants.

\change{\paragraph{Quantitative}}
\change{To analyze this data, we fit four ordinal logistic regression models, one for demographics alone and one for each of our three research question, where the dependent variable was a binary categorical variable representing the group a participant belonged to (0 for unvaccinated, 1 for vaccinated). Our independent variables in these models were our key demographic variables and the influence, agreement, or danger perceptions across the five social circles. These models allowed us to compare the predictive power of perceived influence, agreement or danger relative to demographics and across strong and weak ties. All the independent variables were coded as ordered scales starting at 1. For example, for \textit{Family Influence}, `None at all' is coded as 1, `A little' is coded as 2, `A moderate amount' is coded as 3, `A lot' is coded as 4, `A great deal' is coded as 5. Hence, the Likert scale value increases as \textit{Family Influence} increases.. The coefficient estimates provided are given in units of ordered logits, or ordered log odds. Hence, for a one unit increase in the independent variable, the movement from the unvaccinated group to the vaccinated group is expected to change by its respective regression coefficient in the ordered log-odds scale while the other independent variables in the model are held constant. The ordinal logistic regressions were built using the \textit{statsmodel} (version 0.14.2) library in Python and use the Broyden–Fletcher–Goldfarb–Shanno (BFGS) method for solving.}

\change{\paragraph{Qualitative}}
To analyze our open-ended reasoning question, we take a multi-round, open-coding approach, \change{similar to the approach used in \cite{horne2020tailoring}}. First, two of the authors individually went over the open-ended responses from the vaccinated group and the unvaccinated group to identify broad codes, continuing until the list of codes was stable. The two authors then discuss their codes to develop a combined list of codes, each with specific definitions of what reasoning fits into the category. The final set of codes can be found in Tables \ref{tbl:yes_code_descr} and \ref{tbl:no_code_descr} \change{at the end of the paper.} Then the same two authors coded all of the responses for both groups (vaccinated and not). Since the respondents could describe as many reasons as they wanted, each participants response could fall into more than one category. After the second round of coding, agreement was computed, where a response is said to have agreement if all codes match across the two coders. In the unvaccinated group, 79.84\% of the responses had full agreement, while the vaccinated group, 85.38\% of the responses had full agreement. The two authors met again to discuss and resolve these conflicts. Disagreements in the unvaccinated group were mostly due to interpreting what entity was being distrusted (science, government, media, pharmaceutical companies, general), hence we merged these to a broader category of \texttt{Distrust}. For example, some response were very specific: ``Because there's not enough research to determine if it's safe and I do not trust the government or media.'' while others were more general: ``i dont trust any of it''. Disagreements in the vaccinated group mostly were due to differentiating the categories \texttt{Protect others} versus \texttt{Right thing to do}. In this case, we opted for stricter definitions of each to avoid misinterpretation. For example, a response had to use the phrase ``it was the right thing to do'' to be included in the category \texttt{Right thing to do}, while calls to protecting others, such as ``To protect my family, community'' or ``I wanted to protect others from catching Covid and help reduce the spread'' were put in the category \texttt{Protect others}. Definitions and examples of the final agreed upon results can be found in Tables \ref{tbl:qual_results}, \ref{tbl:yes_code_descr}, and \ref{tbl:no_code_descr}.

Lastly, for our questions on social circle size and trust, we manually parse and count the number of people listed by participants for each question. We do this task manually as participants format these lists in a variety of ways, making automated parsing prone to errors. The distributions of approximate social circle size and social circle trust across the vaccinated and unvaccinated groups are then analyzed quantitatively using a Mann-Whitney test. \\

\section{Results}\label{sec:results}

\begin{figure*}[ht!]
    \centering
    \subfloat[\centering U.S. Counties Where Participants Resided]{{{\includegraphics[trim=0 0cm 0cm 0,clip,width=6.5cm]{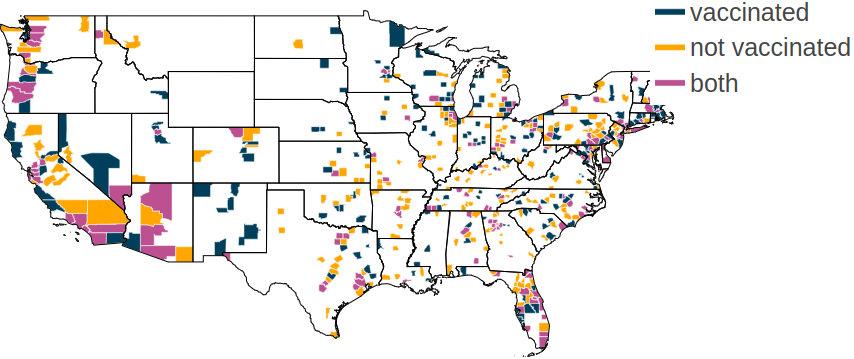} }}}
    \subfloat[\centering Correlation of Demographics]{{{\includegraphics[trim=0 0cm 0cm 0,clip,width=5cm]{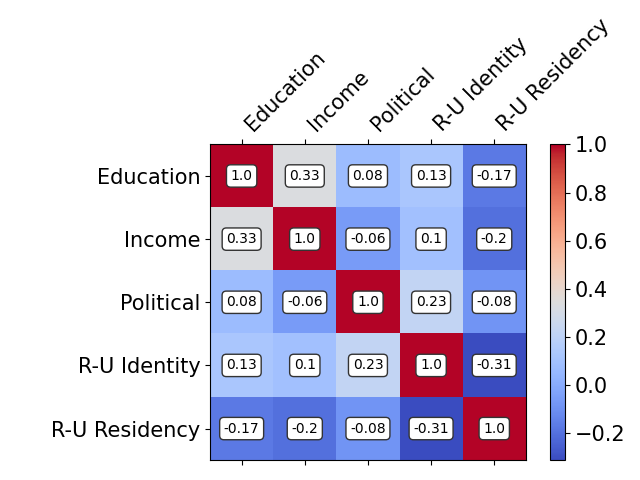} }}}\\
    \subfloat[\centering Political Leaning]{{{\includegraphics[trim=0 0cm 0cm 0,clip,width=3.5cm]{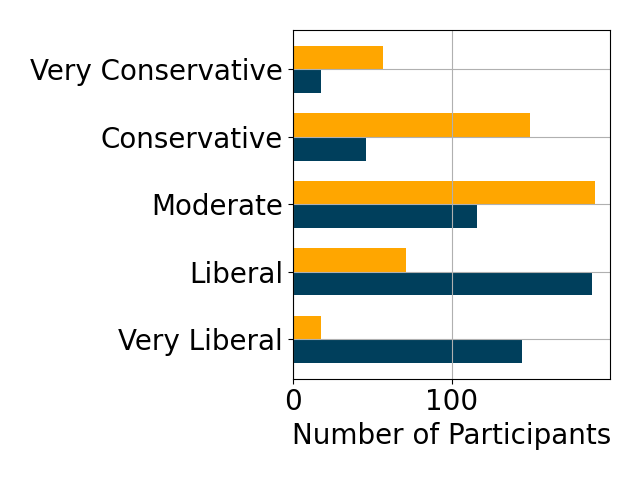} }}}\hfill
    \subfloat[\centering Rural-Urban Identity]{{{\includegraphics[trim=0 0cm 0cm 0,clip,width=3.5cm]{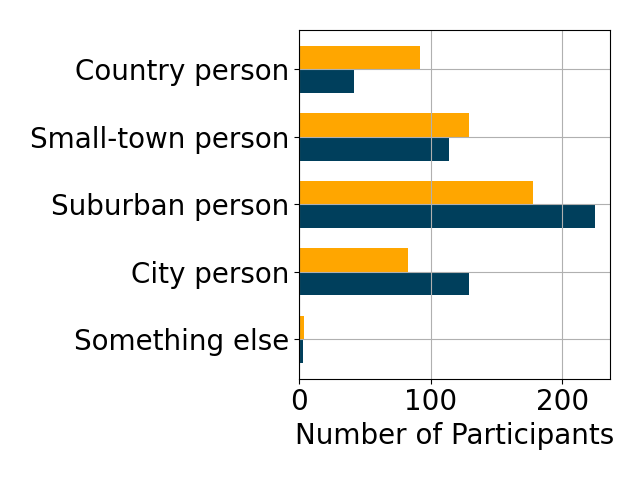} }}}\hfill
    \subfloat[\centering Education]{{{\includegraphics[trim=0 0cm 0cm 0,clip,width=3.5cm]{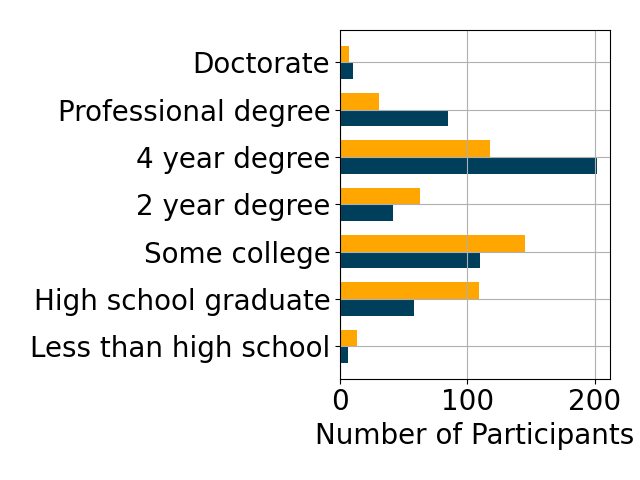} }}}\\
    \includegraphics[width=5cm]{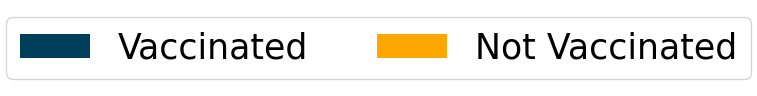}\\
    \caption{\change{(a) Counties where at least one survey participant resides. Participants in our survey lived in 882 unique U.S. counties, with 471 counties represented in the vaccinated group and 461 counties represented in the unvaccinated group. (b) Correlation of key demographic variables used in study. Distributions of (c) political leaning, (d) rural-urban identity, and (e) education of vaccinated (blue) and unvaccinated (yellow) participants.}}%
    \label{fig:fips}%
\end{figure*}

\change{\subsection{Demographic Differences}}\label{sec:results_demo}
\change{Before examining our research questions, we describe the demographic differences across the vaccinated and unvaccinated groups.} In Figure \ref{fig:fips}, \change{we show the correlation among our key demographic variables and} distributions of political leaning, rural-urban identity, and education across vaccinated and unvaccinated groups. Additionally, in Figure \ref{fig:fips}a, we show a map of the U.S. counties in which survey participants resided in at the time of the survey. \change{In Table \ref{tbl:demo}, we show the results of an ordinal logistic regression mdoel, where each demographic question is an independent variable and whether a participant was vaccinated or not is the dependent variable.}

\begin{table}[ht!]
\centering
\begin{tabular}{lcccccc}
\toprule
                            & \textbf{coef} & \textbf{std err} & \textbf{z} & \textbf{P$> |$z$|$} & \textbf{[0.025} & \textbf{0.975]}  \\
\midrule
\textbf{Education}          &       0.3130  &        0.058     &     5.429  &         0.000***        &        0.200    &        0.426     \\
\textbf{Income}             &       0.1218  &        0.025     &     4.963  &         0.000***        &        0.074    &        0.170     \\
\textbf{Political Leaning}   &       1.0445  &        0.080     &    13.126  &         0.000***        &        0.889    &        1.200     \\
\textbf{Rural-Urban Identity} &       0.1310  &        0.085     &     1.534  &         0.125        &       -0.036    &        0.298     \\
\textbf{Rural-Urban Residency}               &       0.0166  &        0.048     &     0.348  &         0.728        &       -0.077    &        0.110     \\
\textbf{0.0/1.0}            &       5.7225  &        0.474     &    12.074  &         0.000***        &        4.794    &        6.651     \\
\bottomrule
\end{tabular}
\caption{\change{Ordinal logistic regression results for key demographic variables. The DV is a binary variable of which group a participant is in, where 1 is for vaccinated and 0 is for unvaccinated. Significance codes are shown for p: *** $p < 0.001$, ** $p < 0.01$, * $p < 0.05$.}}\label{tbl:demo}
\end{table}

Across both the vaccinated and unvaccinated groups, participants described themselves as 40.7\% male, 57.0\% female, and 2.0\% non-binary/third gender, and 86.1\% white, 6.8\% black, 3.5\% Asian, 2.8\% other, and 0.8\% Native American. There were no significant differences found between gender or race across the two groups. There were significant differences across other demographic traits, such as political leaning, income, and education. Specifically, vaccinated participants were significantly more left leaning, had higher income, and were more educated. \change{When examining demographic distributions independently, we found some evidence that there were differences between rural-urban identity and rural residency; namely, vaccinated participants identified more as city/suburban people and lived in more urban areas (shown in the Supplemental materials). However, when controlling for the differences in political leaning, income, and education, these differences become non-significant.}

These demographic differences align with results from previous work. For example, reports from The COVID States project\footnote{\url{https://www.covidstates.org/}} found that people who lived in city/suburban areas, had more education, and had higher income, had significantly higher vaccination rates and greater support for vaccine requirements \cite{lazer2021covid, baum2021covid}. Other studies have shown that rural residents are ``significantly less likely to participate in COVID-19-related preventive health behaviors \cite{callaghan2021rural}'' and that rural residents have higher rates of vaccine resistance \cite{lazer2021covid}. Further, as stated by Green et al. (2022), ``partisanship remain[s] the most stable and sizable gap'' in COVID-19 behaviors and attitudes \cite{green2022using}. Prior work has shown that vaccine hesitancy is significantly higher for those who identify as Republications than those who identify as Democrats \cite{reiter2020acceptability, perlis2020covid, khubchandani2021covid, milligan2021covid, clinton2021partisan}. The demographic results from our survey provide further robustness to these prior findings as well as validate our survey data.

\change{\subsection{Perceived Social Influence (RQ1)}}

\begin{table}[ht!]
\centering
\begin{tabular}{lcccccc}
\toprule
                                   & \textbf{coef} & \textbf{std err} & \textbf{z} & \textbf{P$> |$z$|$} & \textbf{[0.025} & \textbf{0.975]}  \\
\midrule
\textbf{Education}                 &       0.3430  &        0.062     &     5.559  &         0.000***        &        0.222    &        0.464     \\
\textbf{Income}                    &       0.1075  &        0.026     &     4.112  &         0.000***        &        0.056    &        0.159     \\
\textbf{Political Leaning}          &       1.0386  &        0.084     &    12.321  &         0.000***        &        0.873    &        1.204     \\
\textbf{Rural-Urban Identity}        &       0.0966  &        0.089     &     1.089  &         0.276        &       -0.077    &        0.271     \\
\textbf{Rural-Urban Residency}                      &       0.0073  &        0.050     &     0.147  &         0.883        &       -0.091    &        0.106     \\
\textbf{Family Influence}           &       0.3097  &        0.077     &     4.014  &         0.000***        &        0.158    &        0.461     \\
\textbf{Close Friends Influence}     &       0.2184  &        0.104     &     2.095  &         0.036*        &        0.014    &        0.423     \\
\textbf{Coworkers Influence}        &       0.2119  &        0.148     &     1.429  &         0.153        &       -0.079    &        0.502     \\
\textbf{Church/Club Influence} &      -0.3445  &        0.212     &    -1.626  &         0.104        &       -0.760    &        0.071     \\
\textbf{Neighbors Influence}        &       0.4425  &        0.195     &     2.268  &         0.023*        &        0.060    &        0.825     \\
\textbf{0.0/1.0}                   &       7.0613  &        0.576     &    12.253  &         0.000***        &        5.932    &        8.191     \\
\bottomrule
\end{tabular}
\caption{\change{Ordinal logistic regression results for perceived social influence controlling for key demographic variables. The DV is a binary variable of which group a participant is in, where 1 is for vaccinated and 0 is for unvaccinated. Significance codes are shown for p: *** $p < 0.001$, ** $p < 0.01$, * $p < 0.05$.}}\label{ols:influence}
\end{table}

\begin{figure*}[ht!]
    \centering
    \subfloat[\centering Family]{{{\includegraphics[trim=0 0cm 0cm 0,clip,width=3.6cm]{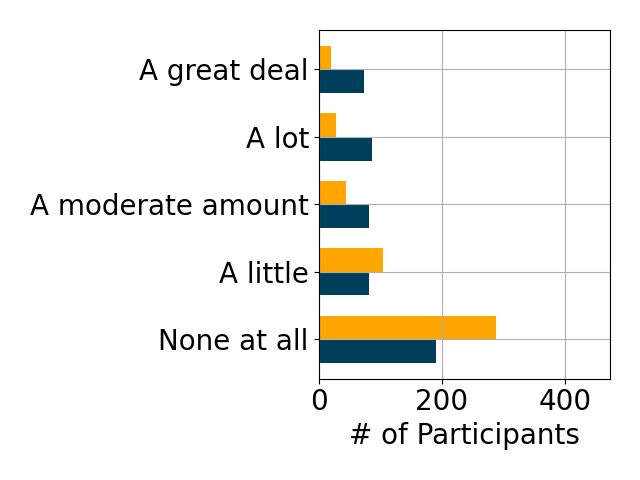} }}}\hfill
    \subfloat[\centering Close Friends]{{{\includegraphics[trim=0 0cm 0cm 0,clip,width=3.6cm]{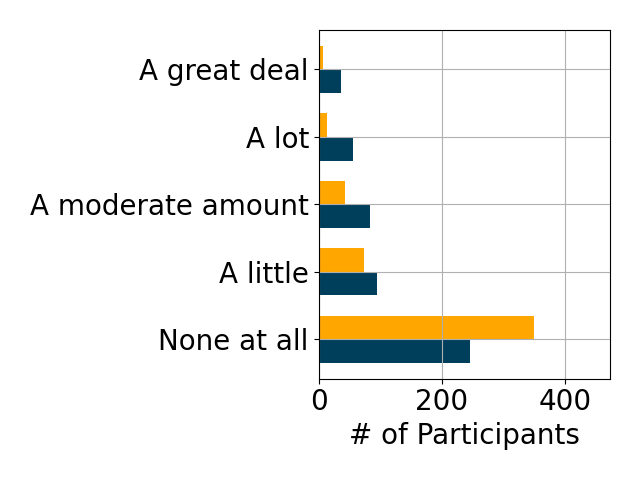} }}}\hfill
    \subfloat[\centering Coworkers]{{{\includegraphics[trim=0 0cm 0cm 0,clip,width=3.6cm]{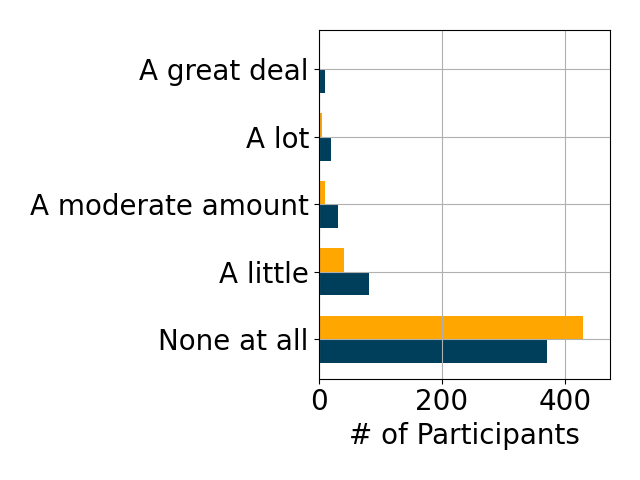} }}}\\
    \subfloat[\centering Church/Social Club]{{{\includegraphics[trim=0 0cm 0cm 0,clip,width=3.6cm]{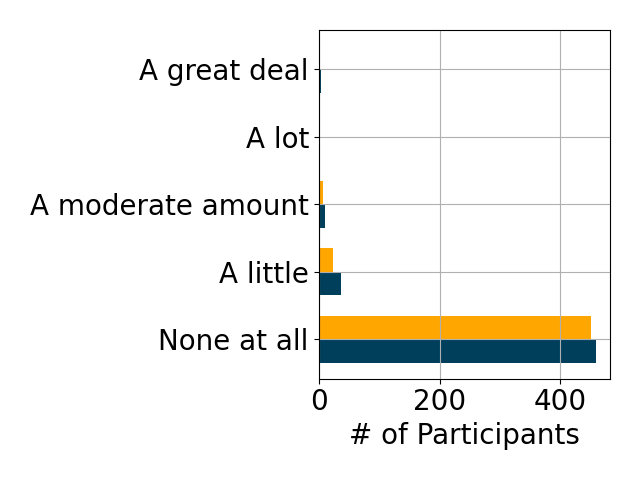} }}}
    \subfloat[\centering Neighbors]{{{\includegraphics[trim=0 0cm 0cm 0,clip,width=3.6cm]{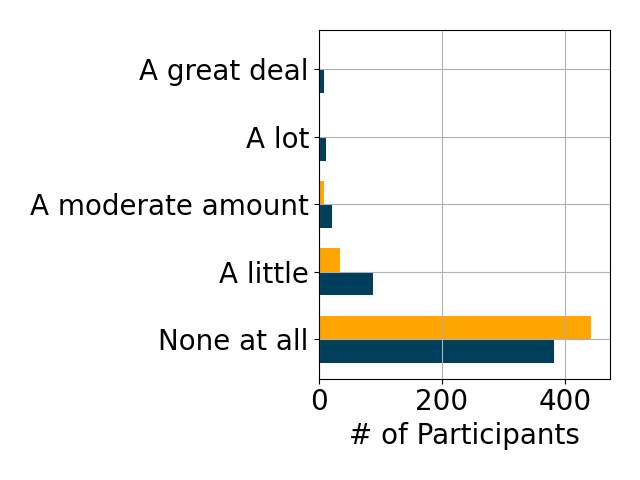}}}}\\
    \includegraphics[width=5cm]{figures/legend.png}\\
    \caption{\change{Distributions of perceived influence across social groups of vaccinated (blue) and unvaccinated (yellow) participants. The means perceived influence across each social group, where `None at all' is 1 and `A great deal' is 5, where: (a) 2.57 for vaccinated, 1.74 for unvaccinated, (b) 2.11 for vaccinated, 1.47 for unvaccinated, (c), 1.47 for vaccinated, 1.17 for unvaccinated, (d) 1.15 for vaccinated, 1.12 for unvaccinated, (e) 1.40 for vaccinated, 1.12 for unvaccinated.}}%
    \label{fig:influence_dists}%
\end{figure*}

First, we examine differences in perceived social influence of theoretically strong and weak social ties across vaccinated and unvaccinated participants. \change{In Table \ref{ols:influence}, we again show the results of fitting an ordinal logistic regression model, where the dependent variable is if the participant is vaccinated or not. However, now we add our influence variables to the model (i.e., ``How much influence did each social group have on your decision to (not) get vaccinated?''). This analysis allows us to show differences in perceived influence relative to other social groups and control for demographic differences. In Figure \ref{fig:influence_dists}, we show the distributions of perceived influence across each social circle in terms of $N$ per Likert scale answer per vaccination group.} 

\change{What is clear from this analysis is that vaccinated participants perceived significantly higher influence from their family than unvaccinated participants ($p < 0.001$). Vaccinated participants also perceived significantly more social influence from close friends and neighbors (each $p < 0.05$), but these differences were less than the perceived influence of family. On average, vaccinated participants fell between `a little' and `a moderate amount' for family influence ($\mu=2.57$, $\sigma=1.48$), between `a little' and `a moderate amount' for close friends influence ($\mu=2.11$, $\sigma=1.30$), and between  `None at all' and `A little' for neighbor influence ($\mu=1.40$, $\sigma=0.81$). On the other hand, unvaccinated participants fell between `None at all' and `A little' for family influence ($\mu=1.74$, $\sigma=1.02$), close friends influence ($\mu=1.46$, $\sigma=0.88$), and neighbor influence ($\mu=1.12$, $\sigma=0.44$). Despite some significant differences across these groups, the majority of participants from both the vaccinated and unvaccinated groups did not perceive any social influence across all social ties (as shown in Figure \ref{fig:influence_dists}).} There were no significant differences in the perceived influence of coworkers or church/club members. When examining the distributions in Figure \ref{fig:influence_dists}, we see generally the same trend across both groups. That is, perceived influence by both groups generally weakens as we move from theoretically stronger relationships to weaker relationships (e.x. family to church/club).

\change{\subsection{Perceived Social Agreement (RQ2)}}

\begin{table}[ht!]
\centering
\begin{tabular}{lcccccc}
\toprule
                                   & \textbf{coef} & \textbf{std err} & \textbf{z} & \textbf{P$> |$z$|$} & \textbf{[0.025} & \textbf{0.975]}  \\
\midrule
\textbf{Education}                 &       0.2442  &        0.070     &     3.494  &         0.000***        &        0.107    &        0.381     \\
\textbf{Income}                    &       0.1174  &        0.030     &     3.974  &         0.000***        &        0.060    &        0.175     \\
\textbf{Political Leaning}          &       1.0520  &        0.094     &    11.216  &         0.000***        &        0.868    &        1.236     \\
\textbf{Rural-Urban Identity}        &       0.0931  &        0.098     &     0.952  &         0.341        &       -0.099    &        0.285     \\
\textbf{Rural-Urban Residency}                      &       0.0372  &        0.054     &     0.694  &         0.488        &       -0.068    &        0.142     \\
\textbf{Family Agree}           &       0.7953  &        0.143     &     5.555  &         0.000***        &        0.515    &        1.076     \\
\textbf{Close Friends Agree}     &       1.0437  &        0.176     &     5.924  &         0.000***        &        0.698    &        1.389     \\
\textbf{Coworkers Agree}        &       0.5954  &        0.202     &     2.951  &         0.003**        &        0.200    &        0.991     \\
\textbf{Church/Club Agree} &      -0.6762  &        0.230     &    -2.937  &         0.003**        &       -1.127    &       -0.225     \\
\textbf{Neighbors Agree}        &       1.2664  &        0.210     &     6.042  &         0.000***        &        0.856    &        1.677     \\
\textbf{0.0/1.0}                   &      12.5360  &        0.838     &    14.953  &         0.000***        &       10.893    &       14.179     \\
\bottomrule
\end{tabular}
\caption{\change{Ordinal logistic regression results for perceived social agreement controlling for key demographic variables. The DV is a binary variable of which group a participant is in, where 1 is for vaccinated and 0 is for unvaccinated. Significance codes are shown for p: *** $p < 0.001$, ** $p < 0.01$, * $p < 0.05$.}}\label{ols:agree}
\end{table}

\begin{figure*}[ht!]
    \centering
    \subfloat[\centering Family]{{{\includegraphics[trim=0 0cm 0cm 0,clip,width=3.6cm]{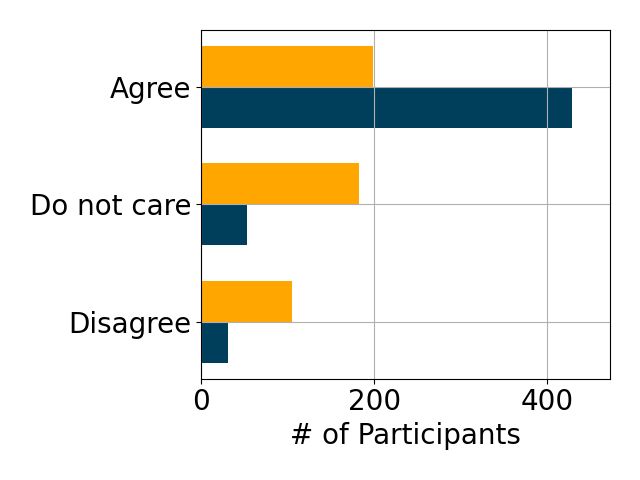} }}}\hfill
    \subfloat[\centering Close Friends]{{{\includegraphics[trim=0 0cm 0cm 0,clip,width=3.6cm]{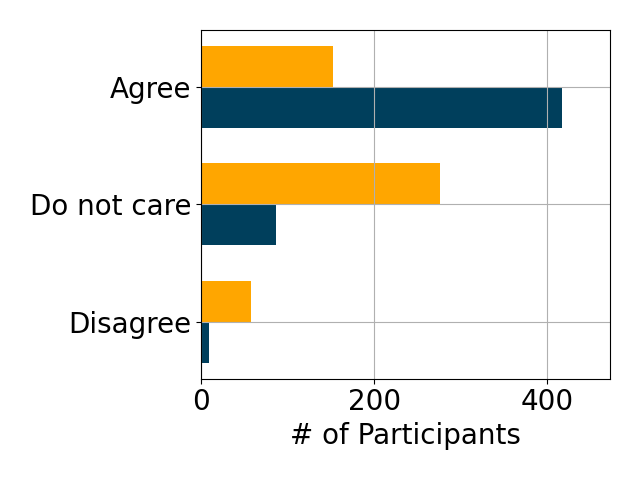} }}}\hfill
    \subfloat[\centering Coworkers]{{{\includegraphics[trim=0 0cm 0cm 0,clip,width=3.6cm]{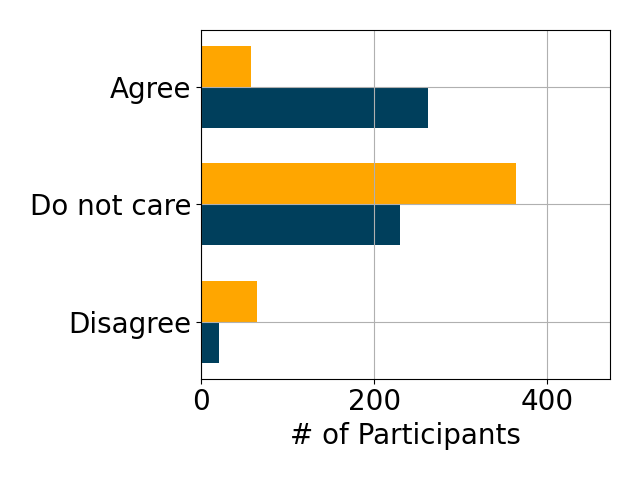} }}}\\
    \subfloat[\centering Church/Social Club]{{{\includegraphics[trim=0 0cm 0cm 0,clip,width=3.6cm]{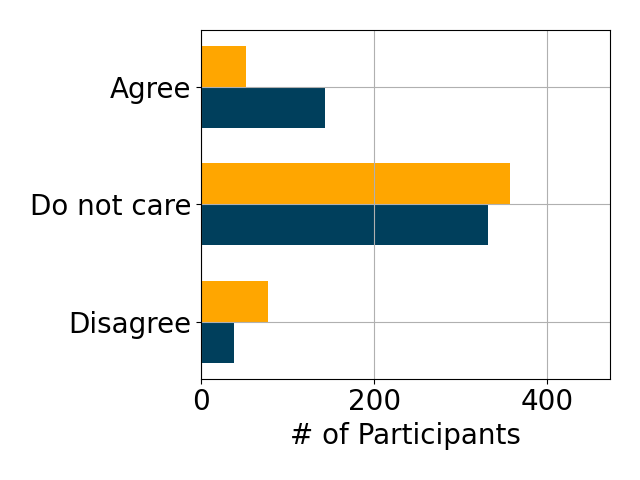} }}}
    \subfloat[\centering Neighbors]{{{\includegraphics[trim=0 0cm 0cm 0,clip,width=3.6cm]{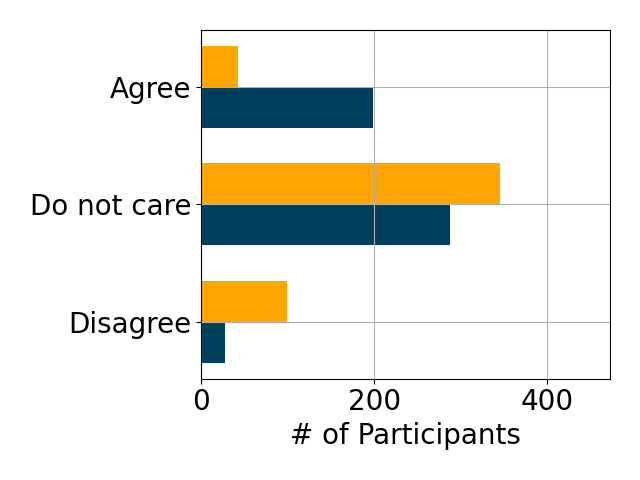}}}}\\
    \includegraphics[width=5cm]{figures/legend.png}\\
    \caption{\change{Distributions of perceived agreement across social groups of vaccinated (blue) and unvaccinated (yellow) participants. The means perceived agreement across each social group, where `Disagree' is 1 and `Agree' is 3, where: (a) 2.78 for vaccinated, 2.19 for unvaccinated, (b) 2.80 for vaccinated, 2.19 for unvaccinated, (c), 2.48 for vaccinated, 1.98 for unvaccinated, (d) 2.21 for vaccinated, 1.95 for unvaccinated, (e) 2.34 for vaccinated, 1.88 for unvaccinated.}}%
    \label{fig:agree_dists}%
\end{figure*}

Next, we examine the differences between vaccinated and unvaccinated participants in perceived agreement from varying social circles. \change{In Table \ref{ols:agree}, we show the results of fitting an ordinal logistic regression model, where the dependent variable is if the participant is vaccinated or not and a mix of demographic controls and perceived agreement are independent variables (i.e., ``How much do you think each social group agrees with your decision to (not) get vaccinated?''). In Figure \ref{fig:agree_dists}, we show the distributions of perceived agreement across varying social circles.}

At a high level, no matter the social group, vaccinated participants perceived significantly higher agreement with others than participants who were not vaccinated. \change{Perceived agreement from family ($p<0.001$), close friends ($p<0.001$), coworkers ($p<0.01$), and neighbors ($p<0.001$) all were perceived as higher for vaccinated participants than unvaccinated participants.} The only exception to this finding is the perceived influence of one’s ‘Church or Social club’. \change{On average, vaccinated participants reported perceived agreement between 'Do not care' and 'Agree' for family ($\mu=2.78$, $\sigma=0.53$), close friends ($\mu=2.80$, $\sigma=0.44$), coworkers ($\mu=2.48$, $\sigma=0.57$), neighbors ($\mu=2.34$, $\sigma=0.57$), and church/club ($\mu=2.20$, $\sigma=0.56$). On average, unvaccinated participants reported perceived agreement between `Do not care' and `Agree' for family ($\mu=2.19$, $\sigma=0.76$) and close friends ($\mu=2.20$, $\sigma=0.63$), although closer to `Do not care' and to 'Agree'. Further, unvaccinated participants perceived agreement between `Disagree' and `Do not care' for coworkers ($\mu=1.99$, $\sigma=0.50$), neighbors ($\mu=1.88$, $\sigma=0.53$), and church/club ($\mu=1.94$, $\sigma=0.52$). The same general trend of weakening as we move from theoretically stronger relationships to weaker relationships that we saw for perceived influence also occurs for perceived agreement, particularly for the vaccinated group.}

\change{\subsection{Perceived Social Danger (RQ3)}}

\begin{table}[ht!]
\centering
\begin{tabular}{lcccccc}
\toprule
                                & \textbf{coef} & \textbf{std err} & \textbf{z} & \textbf{P$> |$z$|$} & \textbf{[0.025} & \textbf{0.975]}  \\
\midrule
\textbf{Education}              &       0.3236  &        0.059     &     5.468  &         0.000***        &        0.208    &        0.440     \\
\textbf{Income}                 &       0.1391  &        0.025     &     5.487  &         0.000***       &        0.089    &        0.189     \\
\textbf{Political Leaning}       &       0.9763  &        0.082     &    11.953  &         0.000***        &        0.816    &        1.136     \\
\textbf{Rural-Urban Identity}     &       0.1153  &        0.086     &     1.334  &         0.182        &       -0.054    &        0.285     \\
\textbf{Rural-Urban Residency}                   &       0.0082  &        0.048     &     0.169  &         0.866        &       -0.087    &        0.103     \\
\textbf{Family Danger}           &       0.1300  &        0.096     &     1.356  &         0.175        &       -0.058    &        0.318     \\
\textbf{Close Friends Danger}     &       0.0447  &        0.119     &     0.377  &         0.706        &       -0.188    &        0.277     \\
\textbf{Coworkers Danger}        &       0.0152  &        0.129     &     0.118  &         0.906        &       -0.237    &        0.268     \\
\textbf{Church/Club Danger} &       0.0923  &        0.125     &     0.740  &         0.459        &       -0.152    &        0.337     \\
\textbf{Neighbors Danger}        &       0.1904  &        0.126     &     1.506  &         0.132        &       -0.057    &        0.438     \\
\textbf{0.0/1.0}                &       6.7926  &        0.543     &    12.510  &         0.000***        &        5.728    &        7.857     \\
\bottomrule
\end{tabular}
\caption{\change{Ordinal logistic regression results for perceived social danger controlling for key demographic variables. The DV is a binary variable of which group a participant is in, where 1 is for vaccinated and 0 is for unvaccinated. Significance codes are shown for p: *** $p < 0.001$, ** $p < 0.01$, * $p < 0.05$.}}\label{ols:danger}
\end{table}

\begin{figure*}[ht!]
    \centering
    \subfloat[\centering Family]{{{\includegraphics[trim=0 0cm 0cm 0,clip,width=3.6cm]{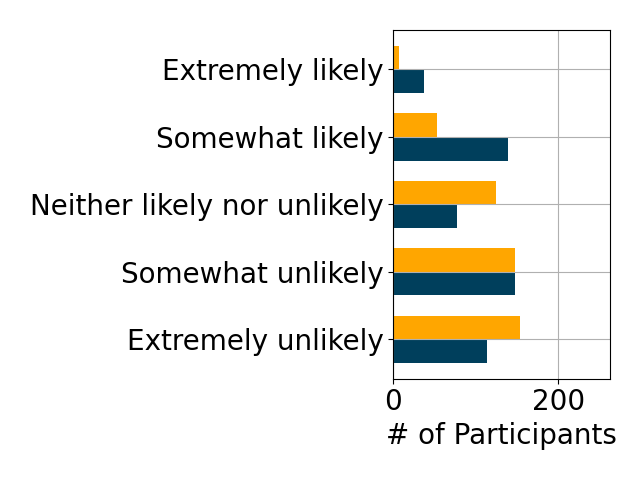} }}}\hfill
    \subfloat[\centering Close Friends]{{{\includegraphics[trim=0 0cm 0cm 0,clip,width=3.6cm]{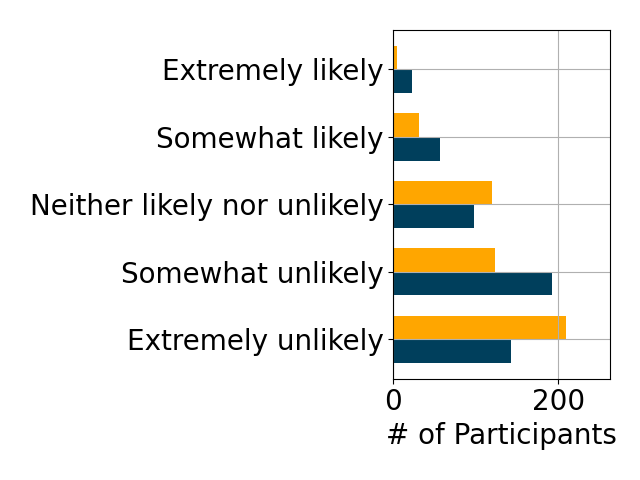} }}}\hfill
    \subfloat[\centering Coworkers]{{{\includegraphics[trim=0 0cm 0cm 0,clip,width=3.6cm]{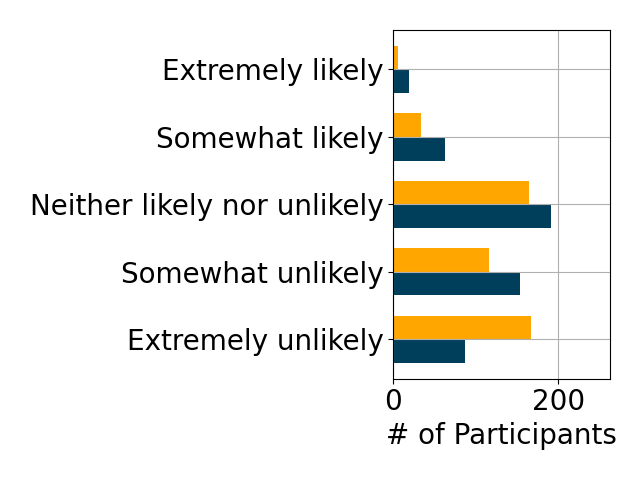} }}}\\
    \subfloat[\centering Church/Social Club]{{{\includegraphics[trim=0 0cm 0cm 0,clip,width=3.6cm]{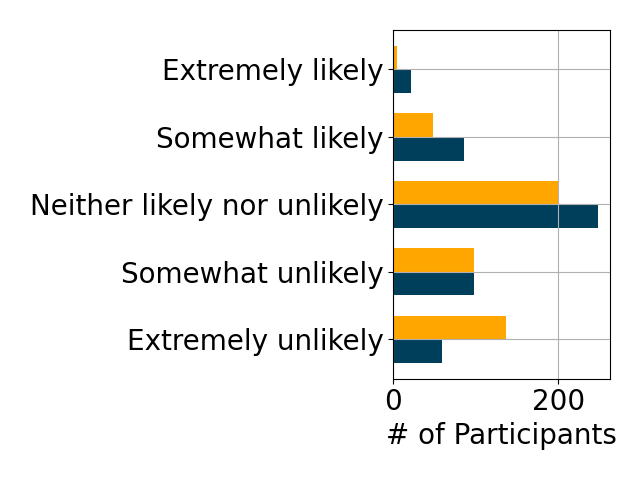} }}}
    \subfloat[\centering Neighbors]{{{\includegraphics[trim=0 0cm 0cm 0,clip,width=3.6cm]{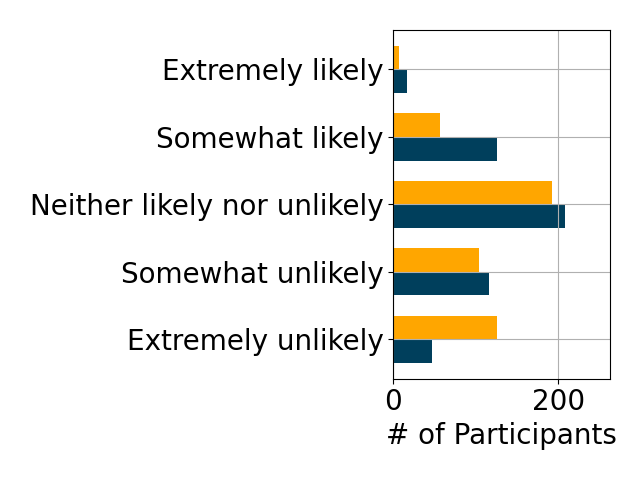}}}}\\
    \includegraphics[width=5cm]{figures/legend.png}\\
    \caption{\change{Distributions of perceived danger across social groups of vaccinated (blue) and unvaccinated (yellow) participants. The means perceived danger across each social group, where `Extremely unlikely' is 1 and `Extremely likely' is 5, where: (a) 2.70 for vaccinated, 2.28 for unvaccinated, (b) 2.28 for vaccinated, 1.97 for unvaccinated, (c), 2.55 for vaccinated, 2.17 for unvaccinated, (d) 2.82 for vaccinated, 2.36 for unvaccinated, (e) 2.90 for vaccinated, 2.42 for unvaccinated.}}%
    \label{fig:danger_dists}%
\end{figure*}

Lastly, we examine how the perceived danger of COVID-19 to varying social groups changes across vaccinated and unvaccinated participants. Specifically, we ask: ``How likely do you think members of each social group are to get severe COVID-19 or die from COVID-19?'' - capturing an indirect social influence rather than a direct influence. It is reasonable to assume that if one believes COVID-19 is dangerous to their family and close friends, they may be more likely to get vaccinated to protect them. \change{In Table \ref{ols:danger}, we again show an ordinal logistic regression model, where the dependent variable is if the participant is vaccinated or not and a mix of demographic controls and perceived danger are independent variables. In Figure \ref{fig:danger_dists}, we show the distributions of perceived danger to varying social groups.}

\change{Across all social groups, there were no significant differences in perceived danger between vaccinated and unvaccinated participants. While when examining the distributions independently (not controlling for other factors), we see some shifts of perceived danger between vaccinated and unvaccinated groups, particularly for family (visually can be seen in Figure \ref{fig:agree_dists}a), these differences do not robustly hold. On average, vaccinated participants perceived the danger of COVID-19 between `Somewhat unlikely' to `Neither likely nor unlikely' for all social groups (means ranging from $1.28$ to $2.90$). Similarly, unvaccinated participants perceived the danger of COVID-19 between `Somewhat unlikely' to `Neither likely nor unlikely' for all social groups except close friends (means ranging from $1.97$ to $2.42$).}

While to the best of our knowledge, perceived danger of COVID-19 to specific social groups has not been studied, \change{our results are somewhat unexpected when compared} with more general findings in the literature. Many factors have been shown to predict COVID-19 risk perceptions, such as personal experience with the virus and hearing about the virus from friends and family \cite{dryhurst2020risk}. Hearing about the virus through friends and family was shown to be the strongest predictor of perceived risk, where risk includes both risk of self and risk of others contracting the virus \cite{dryhurst2020risk}. Other non-social factors have also been shown to predict personal risk to COVID-19, including psychological factors and institutional trust \cite{dryhurst2020risk, yildirim2022factor}. However, many work related to risk perceptions, particularly studies later or post the COVID-19 pandemic, have focused on risk perceptions of the vaccine rather than the virus \cite{lazer2022covid}. Perhaps more importantly, it is likely that perceived danger of COVID-19 changed over time, decreasing throughout the pandemic. Our survey reflects views in mid-2021, over a year after the pandemic had started.

\change{\subsection{Additional Context: Social Circle Size and Trust}}
We can further contextualize our findings by approximating the size of and trust in participants social groups. While theoretically, our survey questions measure differences across strong and weak social ties, the strength of the ties between participants and these groups likely changes across individuals. Here we utilize the answers to the questions asked about voluntary conversations participants had within the past week and the county in which participants lived during the pandemic, as described in the methods section.

We find that on average, participants in the unvaccinated group had conversations with 4.02 people (median 4 people, standard deviation 2.45) in the week prior to the survey and trusted 2.27 of them (median 2 people, standard deviation 1.64). On average, participants in the vaccinated group had conversations with 5.26 people (median 4 people, standard deviation 4.83) in the week prior to the survey and trusted 2.76 of them (median 3 people, standard deviation 2.03). Both groups trusted about 60\% of those they had conversations with in the past week, with 60.91\% trusted in the vaccinated group and 60.60\% trusted in the unvaccinated group. Note, these approximate social circle size numbers align with both pre-pandemic and during-pandemic numbers, which both showed a median close circle size of 4 people \cite{hill2003social, tunccgencc2021social}. 

To better measure differences between the two groups, we perform a Mann-Whitney test on each pair of distributions, finding significant differences between approximate social circle size ($p < 0.001$) and the number of individuals trusted ($p < 0.001$), but no significant difference was found between the proportion of social circle members trusted ($p=0.9807$). Hence, vaccinated participants interacted with slightly more people than unvaccinated, but trust in those people is no different between the groups. This approximations of the number of people participants interacted with prior to the survey suggest that the vaccinated population interacted with more people than the unvaccinated population did. Theoretically, it makes sense that interacting with less people could impact one's perceptions of social agreement with COVID-19 preventive measures and COVID-19 itself. In general, this idea is supported in the literature. For example, when studying vaccine messaging strategies, it was found that ``messages evoking harm reduction and people you know were more effective in counties where the virus is spreading more quickly. \cite{green2021covid}'' 

Saliently, there are likely other confounding factors related to location that may have a significant impact on one's pandemic-related decisions. For example, prior work has demonstrated that local news coverage during 2020 and 2021 was best explained by national trends rather than by local conditions, and the themes of that coverage varied across counties in the U.S. \cite{joseph2022local}. Further, there is evidence that there were urban-rural differences in COVID-19 behaviors (despite those differences not showing robustly in this study), and that those behaviors changed depending on the news produced locally. Specifically, Kim et al. (2020) showed that ``rural residents [were] more likely to engage in social distancing behavior than otherwise similar rural residents if their local news [was] produced in a city that is more impacted by COVID-19'' \cite{kim2020effect}. Other work has shown wide variations in vaccine support across areas in the United States \cite{perlis2020covid}.

\change{\subsection{Additional Context: Open-ended Reasoning}}
\begin{table}[ht!]
\begin{tabular}{cc|cc}
% \fontsize{8.0pt}{9.0pt} % change when formatting is added.
% \selectfont
\toprule
\multicolumn{2}{c}{\textbf{Vaccinated}} & \multicolumn{2}{c}{\textbf{Not Vaccinated}}\\\midrule
\textbf{Category} & \textbf{\% of Responses} & \textbf{Category} & \textbf{\% of Responses}\\\midrule
Protect Self & \textbf{74.27\%} & Fear of Vaccine &\textbf{ 36.65\%} \\
Protect Others & \textbf{54.77\%} & Natural Immunity & 7.41\%\\
Trust in Science & \textbf{14.81\%} & Distrust & 6.24\%\\
Employer Mandate & 2.53\% & Not True Vaccine & \textbf{13.26\%}\\
Return to Normal & 7.41\% & Apathy & 7.02\%\\
Social Pressure & 2.73\% & Restricted Access & 6.63\%\\
Trust in Social Circle & 1.75\% & Low Risk of COVID-19 & \textbf{21.64\%}\\
The Right Thing to Do & 5.85\% & Infringement on Rights & 2.73\%\\
Other & 2.14\% & Conspiracy Theory & 5.26\%\\
& & Other & 2.53\%\\
 \bottomrule\\
\end{tabular}
% \begin{tablenotes}
%       \small
%       \item 
%     \end{tablenotes}
    \caption{Percentage of responses that fit into each category across vaccination groups. Note, responses could use multiple categories of reasoning. Hence, the percentages will not add up to 100\%. For definitions of each category, see Tables \ref{tbl:yes_code_descr} and \ref{tbl:no_code_descr} at the end of the paper.}
    \label{tbl:qual_results}
\end{table}

To add more context our results and capture reasoning that may not have been addressed by our research questions, we ask an open ended question to all participants: ``In your own words, why did you (not) get a COVID-19 vaccine?''. As discussed in Section \ref{sec:qualmethods}, we take an open-coding approach to group responses into categories. The final set of categories and corresponding percentage of responses in the category can be found in Table \ref{tbl:qual_results}. Detailed definitions of each category can be found in Tables \ref{tbl:yes_code_descr} and \ref{tbl:no_code_descr} at the end of this paper.

On average, participants in the unvaccinated group wrote more, producing 140.18 characters on average (maximum 1651 characters, minimum 7 characters) compared to only 89.77 characters on average in the vaccinated group (maximum 760 characters, minimum 12 characters). This often meant that participants in the unvaccinated group were listing many more reasons (or used very complex reasoning) for not getting vaccinated than vaccinated participants did for getting vaccinated. 

%\subsubsection{Unvaccinated reasoning}
The most frequent reason used by unvaccinated participants to not get vaccinated was fear of the vaccine, including fear of the vaccine itself or its side effects (36.65\% of responses). For example:

\begin{quote}
    ``I was concerned about the side effects and cost of possible medical treatment. I've reacted horribly to both the tetanus vaccine and flu vaccine.''
\end{quote}

\begin{quote}
    ``I am worried about potential side effects (mainly about long-term ones which are still currently unknown) and I don't like needles.''
\end{quote}

\begin{quote}
    ``I do not believe the benefits outweigh the risks.''
\end{quote}

\begin{quote}
    ``The vaccine is dangerous and proven not to be 100\% effective against the virus. And COVID is nothing more than the flu.''
\end{quote}

The second most frequent reason expressed was a perceived low risk of catching or becoming seriously ill from COVID-19 (21.64\% of responses), aligning with the results from our ordinal question on the perceived dangers of COVID-19 to others.

Other reasons expressed included: claims that the vaccine was ``not a true vaccine'' (13.26\% of responses), claims of natural immunity (7.41\% of responses), distrust in institutions (6.24\% of responses), and one or more conspiracy theories (5.26\% of responses). As with many of the responses from the unvaccinated participants, multiple of these categories were used together. For example, often there were overlaps between the categories \texttt{Ineffective/Not True Vaccine} and \texttt{Distrust} and \texttt{Conspiracy Theory}; such as: 

\begin{quote}
    ``It's not a vaccine, it's a biologic jab intended to kill people.''
\end{quote}

\begin{quote}
    ``You would have to be a fool to not see it for what it is at this point.  I did not get it because it's poison, intentional... technological... poison. Used to kill and control. And part of a larger agenda to get us used to passports, digital IDs and social credit.''
\end{quote}

\begin{quote}
    ``I do not trust the efficacy of the mRNA technology. I do not trust pharmaceutical companies that attempt to hide the results of trials for decades and have been granted total immunity from prosecution. I believe that adverse reactions have been underreported. I think the silencing and censoring of certain immunologists and virologists that have doubts about mRNA, and a substantial contingent of the medical community advocating for lockdowns and isolation except for massive gatherings for "racial justice" have seriously weakened my level of trust in the scientific and medical communities as a whole. I combine this with the illogical insistence that "masks work" despite overwhelming evidence that any type of face covering beyond type n-95 is ineffective.''
\end{quote}

\begin{quote}
    ``They are not real vaccines yet.  Takes a few years.  The side effects and deaths are extreme with many being hidden.  Once one is made that actually is an immunology then I will look into it further.  Right now it is a jab of chemicals.''
\end{quote}

\begin{quote}
    ``They are all under tested and mRNA vaccines were already deemed not safe for human trial due to mass death and vaccine indused autoimmune immune deficiency in all animal test groups. Good luck with that.''
\end{quote}

Some of these findings align with prior work. For example, research from The COVID States project demonstrated that ``the biggest expressed concern of the unvaccinated is the safety of the COVID-19 vaccines'' and that the unvaccinated population is ``more likely to be skeptical of the efficacy and safety of vaccines'' \cite{uslu2021covid}. Further, as others have argued, misinformation and conspiracy theories can play a role in COVID-19 decision making \cite{pierri2022online}. 

Overall, vaccinated participants expressed very different reasons for getting vaccinated than unvaccinated participants expressed for not getting vaccinated. The vast majority of vaccinated participants stated that they chose to get vaccinated to protect themselves and protect others (74.27\% of responses contained protect self and 54.77\% contained protect others), and typically used very short, simple reasoning. For example, the majority of the responses looked like the following:

\begin{quote}
    ``To protect myself and my family, community.''
\end{quote}

\begin{quote}
    ``To protect myself and others''
\end{quote}

Other less frequent reasons by vaccinated participants included trust in science (14.81\% of responses), desire to return to normal (7.41\% of responses), and getting vaccinated being ``the right thing to do'' (5.85\% of responses). Notably, participants explicitly stating they got vaccinated due to trusting in their social circle, either through direct advice from social circle, a request to get vaccinated from social circle, or family members also getting vaccinated, only was present in 1.75\% of responses. Further, social pressure, both directly from social circles or indirectly from social norms, was mentioned in only 2.73\% of responses. Hence, while perceived social influence and agreement was high amongst vaccinated participants, it was very likely not the primary reasoning behind the decision \change{(as already suggested by Figures \ref{fig:influence_dists} and \ref{fig:agree_dists})}. 

\section{Discussion}
In this study, we found that vaccinated participants perceived more influence from and agreement with others than unvaccinated participants, particularly from strong social ties like one's family and close friends. These findings broadly align with the findings of previous work. Tun{\c{c}}gen{\c{c}} et al. (2021) showed that the perceived approval and adherence of others to social distancing during COVID-19 predicted participant's adherence to social distancing, particularly when others within one's close social circle approved and adhered to pandemic guidelines \cite{tunccgencc2021social}. However, this study did not demonstrate differences in perceived social influence between groups, rather the authors demonstrated that there is a relationship between individual adherence and perceptions of social circle adherence. We add nuance to this finding, by showing perceptions of influence and approval may not be symmetrically across groups (those who follow pandemic guidelines and those who do not). \change{Our finding of vaccination influences primarily lying within the family is consistent with a study of HPV vaccinations among adolescent girls, whose strongest influence was the mother \cite{Karafillakis2022ITT}.}

In contrast, social influence and agreement does not seem to be the primary factor in deciding not to be vaccinated. In open-ended responses, unvaccinated participants most frequently cited fear of the vaccine and distrust in the institutions who make and support the vaccines. While the primary goal of this study was to examine social influence, it is clear that social influence is not the only mechanism of impression in decision making. Other factors not directly measured in this study that likely also play a role in vaccination decisions include media consumption \cite{schulhofer2013newspapers, hayes2015local, gentzkow2011effect, joseph2022local} and individual, generational, and cultural memory \cite{loftus2005planting, lesthaeghe2014second, ruck2020cultural, horne2023generational}. Our qualitative results support this notion. 

Notably, despite vaccinated participants citing social influence at varying levels in their open-ended responses, they too were motivated by fear, frequently citing that they got vaccinated to protect themselves or to avoid dying from COVID-19. Fear and anxiety being a driver of health-related decision making is well-supported in the literature \cite{turk2006using, devlin2007comparative, krishen2015fear}, including behaviors during the COVID-19 pandemic. For example, Harper et al. (2021) found that from a sample of 324 people from the U.K. that ``the only predictor of positive behavior change (e.g., social distancing, improved hand hygiene) was fear of COVID-19 \cite{harper2021functional}'', including when controlling for political and ideological variables. When fear and anxiety motivates vaccination, this fear is functional (as Harper et al. (2021) puts it), but, as our study suggests, when fear and anxiety motivates not getting vaccinated, the fear may be counterproductive and even detrimental. 

In conclusion, we found that vaccinated populations perceived more influence from their social circles than unvaccinated populations. This finding held true across various social groups, including family, close friends, and neighbors. Similarly, we found that vaccinated participants perceived that others agreed with their decision to get vaccinated significantly more than unvaccinated participants perceived that others agreed with their decision to not get vaccinated. This finding held across family, close friends, coworkers and neighbors. Indirect measures of social influence \change{did not} follow this trend. Vaccinated participants did not perceive COVID-19 as any more dangerous to their social circles than unvaccinated participants. Despite the clear differences in perceived social influence and agreement across the groups, we found through open-ended responses that both vaccinated and unvaccinated participants frequently cited fear as a motivating factor in their decision: vaccinated participants feared COVID-19, while unvaccinated participants feared the vaccine itself. Together, our results expand the current literature on vaccination behavior and add robustness to several previous findings.

\section{Limitations}
While we are confident in the results presented in this study, it is not without limitations. First, our work is limited in its reliance on self-reported/perceived effects - it is unlikely that perceived influence perfectly correlates with actual influence, which may happen passively or unconsciously. A report from Altay et al. (2022) shows a clear example of this effect, stating that ``[c]onspiracy believers were more likely to report relying less on social information than actually relying less on social information in the behavioral tasks'' \cite{altay2022conspiracy}. Our focus in this work is not on conspiracy believers; however, the general premise of reporting the use of social information versus actual reliance on social information still applies. From this view, it may be that both vaccinated and unvaccinated populations rely on social information, but unvaccinated populations report using less of it than vaccinated populations.

Second, as a whole, the survey data are not from a representative sample of the U.S. population, as clearly shown by the racial demographics in Section \ref{sec:results}. This bias is due to our method of balanced sampling across the vaccinated and unvaccinated groups. Systematically sampling this imbalanced variation allowed us to capture both vaccinated and unvaccinated people's perceptions fairly. While this survey could be re-deployed on a truly representative sample the U.S. population, there would be an imbalance in vaccinated versus unvaccinated participants, as the majority of the U.S. population is vaccinated. Despite this limitation, our survey did cover a broad set of counties and demographics.

\change{Third, our open-ended results suggest fear as a major factor in both getting vaccinated and not getting vaccinated. However, given the study was not designed to measure fear rather social influence and agreement, we cannot robustly establish this result outside of our qualitative analysis.} 

\begin{table}[ht!]
\fontsize{8pt}{9pt}\selectfont
\begin{tabular}{c p{7.4cm}}
\toprule
\multicolumn{2}{c}{\textbf{Vaccinated}}\\\toprule
\textbf{Category} & \textbf{Definition} \\\midrule
Protect Self & Protecting self, to avoid bad illness, or to avoid death\\\midrule
Protect Others & Protecting others, including family, friends, community, neighbors, or ``others''\\\midrule
Trust in Science & Specific mentions of trust in the science or vaccines\\\midrule
Employer Mandate & Specific mentions of employer mandate or job prospects\\\midrule
Desire to Return to Normal & Specific mentions of ``return to normal''\\\midrule
Social Pressure & Various forms of social pressure, such as specific mentions of social pressure or indirect statements like: ``not to look like an anti-vaxxer'' or ``not to be socially unacceptable''\\\midrule
Trust in Social Circle & Conversations with family or friends, family member asking them to get vaccinated, family member also getting vaccinated, etc.\\\midrule
The Right Thing to Do & Specific mentions of getting vaccinated being ``right thing to do''\\\midrule
Other & Response doesn't fit into above categories\\
 \bottomrule\\
\end{tabular}
    \caption{Qualitative Code Book definitions for the vaccinated group}
    \label{tbl:yes_code_descr}
\end{table}

\begin{table}[ht!]
\fontsize{8pt}{9pt}\selectfont
\begin{tabular}{c p{7.4cm}}
\toprule
\multicolumn{2}{c}{\textbf{Not Vaccinated}}\\\toprule
\textbf{Category} & \textbf{Definition} \\\midrule
Fear of Vaccine & Fear of the vaccine itself or the side effects from it. Some mention being afraid of dying from the vaccine, such as ``There have been severe side effects and death after the vaccine.''\\\midrule
Natural Immunity & Specific mentions of having ``natural immunity'' including because of having COVID-19 already, because they take supplements, because ``God'' gave them an ``immune system'', or belief that they ``can fight COVID-19 off naturally'' \\\midrule
Not True Vaccine & Statements of two general forms: 1. The vaccine is ineffective because people can get COVID-19 after being vaccinated, or 2. The belief that the vaccine is "not a true vaccine" for reasons related to ``mRNA'' not being a ``true vaccine'' or because it doesn't always prevent catching COVID-19\\\midrule
Distrust & Specific mentions of distrusting science, the media, the government, ``big pharma'', or ``it''\\\midrule
Apathy & Did not care to get it or did not have time\\\midrule
Restricted Access & Participants have the intention to get the vaccine, but cannot due to health issues, lack of transportation,  or family member not allowing them to get vaccinated\\\midrule
Low Risk of COVID-19 & Participant believes they are not at risk of getting COVID-19 or getting severely ill from COVID-19 \\\midrule
Infringement on Rights & Participant did not get vaccinated because they do not like ``being forced'' to get vaccinated or out of the principle that ``mandates are unacceptable in a free and democratic society'' \\\midrule
Conspiracy Theory & Reasoning using one or more conspiracy theories, where there is a secret, over-arching, malicious plot by powerful groups of people against those not in power, many are framed around blaming democrats, general government, or the media\\\midrule
Other & Response doesn't fit into above categories\\
 \bottomrule\\
\end{tabular}
    \caption{Qualitative Code Book definitions for the unvaccinated group}
    \label{tbl:no_code_descr}
\end{table}

\small{
\noindent \textbf{Funding:} Participants in this study were paid through internal funds in the School of Information Sciences at UTK. No external funding was used for this study.
\\

\noindent \textbf{Ethical Approval:} All research was performed in accordance with relevant guidelines/regulations applicable when human participants are involved (Declaration of Helsinki). The research was approved by The University of Tennessee-Knoxville Human Research Protections Program (HRPP), which determined that the application was eligible for exempt status under 45 CFR 46.104.d, Category 2 ({\footnotesize \url{https://www.hhs.gov/ohrp/regulations-and-policy/regulations/45-cfr-46/revised-common-rule-regulatory-text/index.html#46.101}}). Our application was determined to comply with proper consideration for the rights and welfare of human subjects and the regulatory requirements for the protection of human subjects.
\\

\noindent\textbf{Informed consent:} As per the IRB application, informed consent was acquired from all participants in this survey. All participants were adults living the United States.
\\

\noindent \textbf{Author's contribution:} DY: original idea, survey design, qualitative analysis. RAB: writing, literature review. BDH: survey design, quantitative analysis, qualitative analysis, writing, literature review.
\\

\noindent \textbf{Conflict of interest:} The authors have no competing interests related to this work.
\\

\noindent \textbf{Data availability statement:} Due to the sensitive nature of survey data, the data used in this study is only available upon request.
\\

\noindent \textbf{Supplemental Materials:} Additional survey description and analysis can be found in the supplement.

\bibliography{ref}

\end{document}